%% file: main.tex
\documentclass[conference]{IEEEtran}
\IEEEoverridecommandlockouts
\usepackage{cite}
\usepackage{amsmath,amssymb,amsfonts}
\usepackage{algorithmic}
\usepackage{graphicx}
\usepackage{textcomp}
\usepackage{xcolor}
\def\BibTeX{{\rm B\kern-.05em{\sc i\kern-.025em b}\kern-.08em
    T\kern-.1667em\lower.7ex\hbox{E}\kern-.125emX}}
\begin{document}

\title{Performance Analysis of Deep Learning Models for Femur Segmentation in MRI Scan \\
}



\author{Mengyuan Liu, Yixiao Chen, Anning Tian, Xinmeng Wu, Mozhi Shen, Tianchou Gong, Jeongkyu Lee
\thanks{M. Liu, Y. Chen, A. Tian, X. Wu, M. Shen, T. Gong, J. Lee are with the Khoury College of Computer Sciences, Northeastern University, San Jose, CA, 95113, USA. Email: \{liu.mengyu, chen.yixiao, tian.ann, wu.xinm, shen.moz, gong.tian, jeo.lee\}@northeastern.edu.}%
}

\maketitle

\begin{abstract}
Convolutional neural networks like U-Net excel in medical image segmentation, while attention mechanisms and KAN enhance feature extraction. Meta's SAM 2 uses Vision Transformers for prompt-based segmentation without fine-tuning. However, biases in these models impact generalization with limited data. In this study, we systematically evaluate and compare the performance of three CNN-based models, i.e., U-Net, Attention U-Net, and U-KAN, and one transformer-based model, i.e., SAM 2 for segmenting femur bone structures in MRI scan. The dataset comprises 11,164 MRI scans with detailed annotations of femoral regions. Performance is assessed using the Dice Similarity Coefficient, which ranges from 0.932 to 0.954. Attention U-Net achieves the highest overall scores, while U-KAN demonstrated superior performance in anatomical regions with a smaller region of interest, leveraging its enhanced learning capacity to improve segmentation accuracy.
\end{abstract}


\input{intro}
\input{related_work}

\section{Materials and Methods}

In this section, we present the image dataset and detail the preprocessing steps applied uniformly across all models. Subsequently, we introduce the four deep learning models employed in this study, providing an overview of their architectural and operational characteristics. 

\subsection{Image Dataset and Pre-processing}

The dataset that is utilized in this study comprises lower-body MRI scan from 10 typically developing (TD) children, designated as TD01 through TD10. In total, 11,164 PNG images are employed for analysis.

All images and masks are centralized, resized, and lightly cropped from 546×158 to 360×160 pixels, removing extraneous background while preserving key anatomy for consistent and efficient deep-learning segmentation across 16 datasets from 10 patients.



\subsection{Deep Learning Models for Comparison}

We compare four models: U-Net, Att U-Net, U-KAN, and SAM 2. U-Net and SAM 2 utilize transfer learning \cite{weiss2016survey} to adapt pre-trained architectures for MRI segmentation, leveraging their proven effectiveness in medical imaging.
\begin{itemize}

\item U-Net: Chosen for its ability to handle image noise, blurred boundaries, and limited labeled data, U-Net’s skip connections enhance feature integration, ensuring precise anatomical segmentation.

\item Att U-Net: Incorporates attention mechanisms to dynamically focus on relevant features, improving segmentation accuracy and capturing global dependencies within MRI scans.

\item U-KAN: Enhances U-Net with Kolmogorov-Arnold Networks (KAN), introducing non-linear learnable activation functions to refine accuracy and interpretability in MRI segmentation.

\item SAM 2: A transformer-based model adapted for MRI segmentation via transfer learning. It requires prompts for segmentation predictions, addressed by a random point selection strategy for effective MRI processing.
\end{itemize}

\section{Experimental Setup}
In this section, we describe the unified training framework implemented for all deep learning models to ensure consistency and comparability across methodologies. We also introduce the ensemble strategy designed to approximate the average performance typically achievable from the training process. Finally, we detail the performance metrics used to evaluate and compare the effectiveness of the approaches.

\subsection{Training Setup}
To establish a consistent and robust pipeline for training and evaluating all models in this study, we adopt a unified training approach that incorporates multiple components to enhance segmentation accuracy. The optimization process utilizes a weighted combination of Dice loss (90\%) and boundary loss (10\%), a configuration validated as highly effective for femur bone segmentation \cite{Liu2025}. Sigmoid activation is employed in the output layer, appropriate for the binary nature of the segmentation task.

For the training, 30\% of the high-resolution data is utilized for training, and 10\% is reserved for the testing. Within the training dataset, a 90:10 split is implemented to allocate data for training and validation. Given the potential variability in convergence rates between CNN and transformer-based models, all networks are trained for 100 epochs to ensure thorough optimization, with the best-performing weights selected based on validation performance. Each training procedure is repeated five times to account for variability and enhance reliability.

A learning rate of 0.0001 is initialized to promote gradual and stable convergence during training. To preserve critical anatomical details essential for precise segmentation, input images are maintained at their original resolution. This standardized approach ensures comparability across models and supports robust evaluations of their segmentation capabilities.

\subsection{Ensemble Approach}
To further enhance performance and reliability, an ensemble approach is developed. Predictions from five independently trained models are aggregated using a majority voting mechanism. This ensemble strategy helps mitigate the biases and inconsistencies of individual models and provides a predicted mask for final evaluation. This robust methodology ensures consistent and reliable segmentation results across all tested models.

\subsection{Performance Metrics}
For evaluation, we employ the DSC. The DSC quantifies the pixel-wise concordance between a predicted segmentation and the corresponding ground truth. The coefficient is calculated as follows in equation \eqref{eq:dice}: 

\begin{equation}
    \text{DSC}\left(A, B\right) = \frac{2 |A \cap B|}{|A| + |B|}
    \label{eq:dice}
\end{equation}

\noindent
, where \( A \) represents the predicted set of pixels and \( B \) denotes the ground truth. The DSC ranges from 0, indicating no overlap, to 1, signifying perfect overlap. 

\input{results}
\input{conclusion}

\bibliography{references}
\bibliographystyle{IEEEtran}

\end{document}

%% file: intro.tex
\section{Introduction}

Magnetic Resonance Imaging (MRI) provides detailed anatomical imaging without radiation, making it essential for diagnostics and treatment planning \cite{sciarra2011advances}. However, converting MRI scans into precise femur models is challenging due to the labor-intensive and error-prone manual segmentation process \cite{gu2024segmentanybone}. Automated segmentation is crucial for improving personalized diagnostics in orthopedics and rehabilitation.

Developing reliable segmentation algorithms is difficult due to motion blur, and distortions \cite{Zhang2021}. This is exaggerated by bone morphology variability and low-contrast boundaries\cite{Florkow2022magnetic}. While the U-Net model has achieved a Dice Similarity Coefficient (DSC) \cite{Dice1945measures} of 0.91 \cite{Liu2025}, more accurate and efficient models are still needed.


In this study, we systematically evaluate and
compare the performance of three convolutional neural network (CNN) \cite{Long2015}-based models, i.e., U-Net\cite{Ronneberger2015u}, Attention U-Net (Att U-Net)\cite{oktay2018attention}, and U-Kolmogorov-Arnold Network (U-KAN) \cite{li2024ukanmakesstrongbackbone}, and one transformer-based model, i.e., Segment Anything Model 2 (SAM 2) \cite{ravi2024sam2segmentimages} for segmenting femur bone structures in MRI scan, which are promising architectures renowned for their success in medical imaging applications. This study will address current limitations in segmentation precision and robustness. To benchmark their performance, we compare all four models under a unified training and prediction approach. By targeting a segmentation accuracy surpassing the current state of the art and rigorously evaluating these methods on clinically annotated datasets, this study seeks to automate femur segmentation in MRI scan. Ultimately, the findings aim to contribute to improved clinical outcomes in orthopedics and broader medical applications.

%% file: related_work.tex
\section{Related Work}
In this section, we review recent advancements in deep learning models for medical image segmentation, and state-of-the-art femur segmentation from MRI scan. Through the analysis of existing literature, we highlight the critical importance of systematically comparing the performance of different deep learning models for bone segmentation in MRI to advance the field effectively.

\subsection{Deep Learning Models for Medical Image Segmentation}
Medical image segmentation has seen significant advancements with the development of deep learning models, particularly CNNs and their variants. Among these, U-Net and its derivatives have established themselves as foundational architectures due to their effectiveness in capturing spatial and contextual features through encoder-decoder structures and skip connections. Extensions like U-Net++ \cite{zhou2018unet++}, Att U-Net, and U-KAN have further improved segmentation accuracy in various applications, including MRI scan and CT imaging. 


Another promising approach for medical image segmentation is transformer. It offers two primary strategies for implementation. The first involves integrating transformer architectures with U-Net to enhance spatial feature representation, as demonstrated by models like Swin U-Net \cite{cao2022swin} and TransUNet \cite{chen2021transunet}. These hybrid models leverage the strengths of both transformers and CNNs, achieving improved performance in complex segmentation tasks. The second approach utilizes general-purpose transformer-based models, such as SAM 2, which have broadened the scope of image segmentation. 



\subsection{Femur Segmentation}
Despite advancements, deep-learning models for femur segmentation remain clinically inadequate. The U-Net model achieves a DSC of 0.91\cite{Liu2025}, while the transformer-based SegmentAnyBone model \cite{gu2024segmentanybone}, designed for bone segmentation, attains only 0.72.

\begin{figure}[t]
\centerline{\includegraphics[scale=0.55]{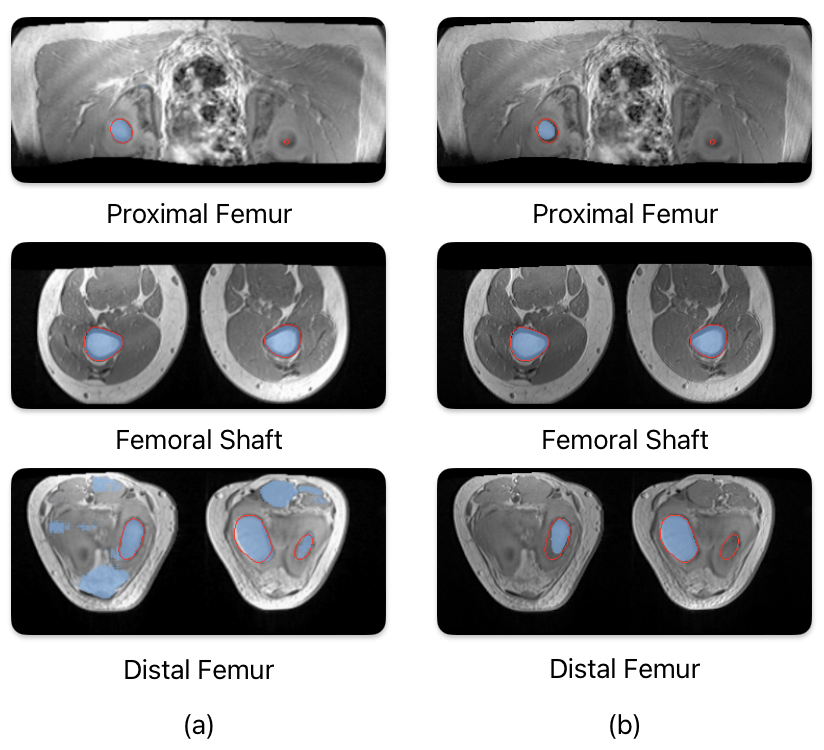}}
\caption{The figure presents segmentation results, with column (a) showing predictions from the SegmentAnyBone model and column (b) displaying results from the U-Net model. Predicted segmentations are marked in blue, while red boundaries denote ground-truth annotations. For SegmentAnyBone, the DSCs are 0.82, 0.92, and 0.56 from top to bottom. For U-Net, the corresponding DSCs are 0.69, 0.97, and 0.82.}
\label{fig: seg+unet}
\end{figure}

As illustrated in Figure \ref{fig: seg+unet} (a), with the assistance of bounding boxes, while the transformer model performs reasonably well on the femoral shaft, its accuracy diminishes significantly on the proximal (hip-side) and distal (knee-side) regions of the femur. Conversely, Figure \ref{fig: seg+unet} (b) shows that while the CNN-based U-Net model demonstrates superior performance on the femoral shaft and distal femur, it still falls short in achieving satisfactory accuracy at the proximal and distal extremities.

This study aims to systematically evaluate the performance of various deep learning models for femur segmentation, with the goal of identifying the most effective approach. By addressing these limitations, this research seeks to advance medical imaging, ultimately enhancing clinical and biomechanical applications.

%% file: results.tex
\section{Experimental Results and Analysis}

The performance of the four models is summarized in Table \ref{table: allMeanSD} based on the DSC. Among the models, Att U-Net achieves the highest mean DSC of 0.954 ± 0.065, highlighting its superior segmentation accuracy and consistency. U-KAN follows closely with a DSC of 0.949 ± 0.090. SAM 2 also delivers strong performance with a high DSC of 0.950 ± 0.035.

\begin{table}[t]
    \centering
    \caption{Comparison of segmentation performance for femur segmentations}
    \renewcommand{\arraystretch}{1.25}
    \begin{tabular}{|c|c|}
    \hline
    \textbf{Models}      & \textbf{DSC (mean ± std)} \\ \hline
    \fontsize{8pt}{11pt}\selectfont
    \textbf{U-Net}       & 0.932 $\pm$ 0.066 \\ \hline
    \fontsize{8pt}{11pt}\selectfont
    \textbf{Att U-Net}   & 0.954 $\pm$ 0.065 \\ \hline
    \fontsize{8pt}{11pt}\selectfont
    \textbf{U-KAN}       & 0.949 $\pm$ 0.090 \\ \hline
    \fontsize{8pt}{11pt}\selectfont
    \textbf{SAM 2}       & 0.950 $\pm$ 0.035 \\ \hline
    \end{tabular}
    \vspace{-5mm}
    \label{table: allMeanSD}
\end{table}



For a more granular analysis, the prediction results are stratified into three distinct anatomical regions. Specifically, the proximal region of the femur, encompassing approximately 20\% of the testing dataset, the femoral shaft region, which constitutes around 60\% of the testing dataset, and the distal region, accounting for the remaining 20\% of the testing dataset.

\begin{figure}[htbp]
\vspace{-3mm}
\centerline{\includegraphics[scale=0.60]{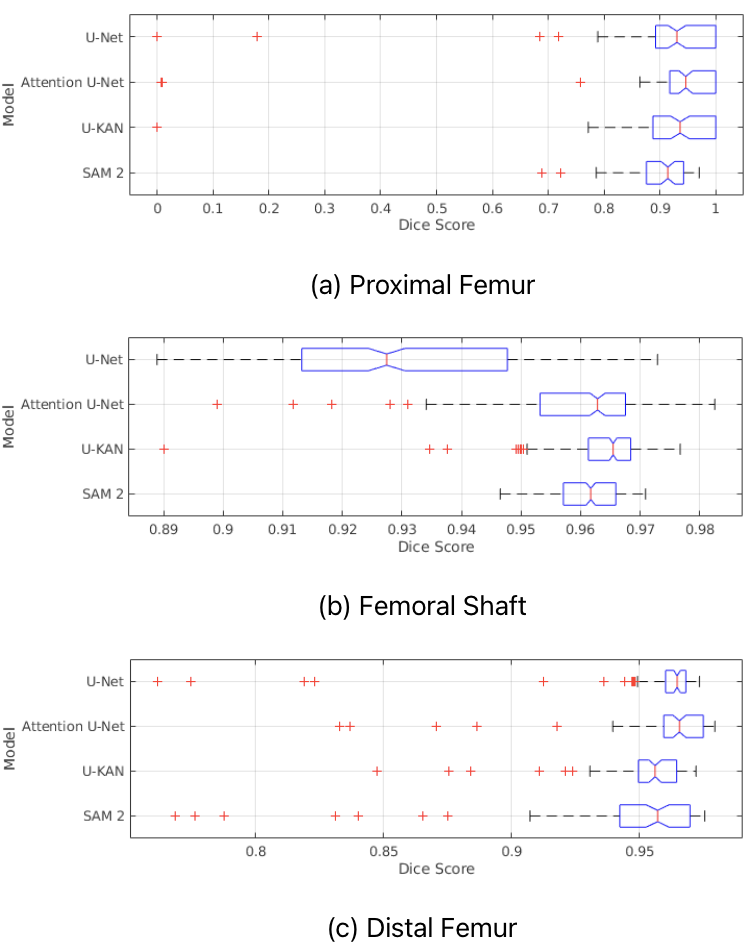}}
\caption{Deep Learning Models Comparison for Different Parts.}
\vspace{-3mm}
\label{fig: partAnalysis}
\end{figure}

Table \ref{table: meanSD} provides a summary of the performance metrics for the four segmentation approaches applied to the three regions of the femur. Corresponding box plots of the DSC are presented in Figure \ref{fig: partAnalysis}.

From Table \ref{table: meanSD}, it is evident that all models achieve a DSC exceeding 0.90 across all anatomical regions, with the Att U-Net demonstrating superior performance compared to the others. This enhanced performance can be attributed to the integration of attention mechanisms, which employ attention blocks to selectively emphasize relevant features while mitigating the influence of irrelevant background noise \cite{oktay2018attention}. These mechanisms facilitate improved feature representation and yield more consistent segmentation results. Such attributes are particularly beneficial in addressing the inherent class imbalance often encountered in medical imaging datasets \cite{Yeung2022unified}.

U-KAN demonstrates the highest segmentation performance in the femoral shaft region, characterized by an exceedingly small region of interest (ROI), as detailed in Table \ref{table: roi}. This is substantiated by its superior median, first-quartile, and third-quartile DSC values, depicted in Figure \ref{fig: partAnalysis} (b). These results suggest that, with adequate training data, U-KAN's advanced learning capacity enabled by the integration of additional learnable nonlinear layers \cite{li2024u} supports highly effective segmentation outcomes. Conversely, U-KAN exhibits reduced performance in the proximal and distal regions, which is likely due to insufficient training data in these areas, leading to a predisposition toward overfitting.

As observed in Figure \ref{fig: partAnalysis} (a), the transformer-based method (SAM 2) demonstrates lower median, first-quartile, and third-quartile DSC values when compared to the U-Net variants. However, SAM 2 exhibits fewer extreme outliers in the DSC range of 0 to 0.2. This behavior can be attributed to SAM 2’s dependency on point or box prompts for its predictions, which prevents it from generating outputs when no bone structures are present in the input image. In contrast, U-Net variants are more prone to false positive predictions in figures without bone, particularly on the proximal region, thereby contributing to the occurrence of lower DSC outliers.

\begin{table}[t]
    \centering
    \caption{Comparison of segmentation performance for different parts femur segmentations. Values are given in mean ± standard deviation format}
    \renewcommand{\arraystretch}{1.25}
    \begin{tabular}{|c|c|c|c|}
    \hline
    \textbf{Models}      & \textbf{Proximal Femur} & \textbf{Femoral Shaft} & \textbf{Distal Femur} \\ \hline
    \fontsize{8pt}{11pt}\selectfont
    \textbf{U-Net} & 0.917 $\pm$ 0.130 & 0.931 $\pm$ 0.022 & 0.956 $\pm$ 0.036 \\ \hline
    \fontsize{8pt}{11pt}\selectfont
    \textbf{Att U-Net} & 0.931 $\pm$ 0.134 & 0.961 $\pm$ 0.012 & 0.961 $\pm$ 0.025 \\ \hline
    \fontsize{8pt}{11pt}\selectfont
    \textbf{U-KAN} & 0.903 $\pm$ 0.185 & 0.964 $\pm$ 0.007 & 0.953 $\pm$ 0.019 \\ \hline
    \fontsize{8pt}{11pt}\selectfont
    \textbf{SAM 2} & 0.904 $\pm$ 0.055 & 0.961 $\pm$ 0.005 & 0.946 $\pm$ 0.041 \\ \hline
    \end{tabular}
    \label{table: meanSD}
\end{table}

\begin{table}[t]
    \centering
    \caption{Different Femur Parts' Average ROI Percentage}
    \renewcommand{\arraystretch}{1.25}
    \begin{tabular}{|c|c|c|c|}
    \hline
    \textbf{Percentage}  & \textbf{Proximal Femur} & \textbf{Femoral Shaft} & \textbf{Distal Femur} \\ \hline
    \fontsize{8pt}{11pt}\selectfont \textbf{ROI (\%)}  & 4.154    &  2.837   & 8.790     \\ \hline
    \end{tabular}
    \vspace{-5mm}
    \label{table: roi}
\end{table}

From Figure \ref{fig: partAnalysis} (b), it is evident that the basic U-Net model exhibits significant variability in predictions, as indicated by the large interquartile range (distance between the first and third quartiles) of the DSC values. Furthermore, the U-Net demonstrates lower median, first-quartile, and third-quartile DSC values compared to the other models, indicating reduced stability and accuracy in body-region predictions. These findings corroborate the observations of \cite{tang20243d}, which highlight the limitations of U-Net models in medical image segmentation. Specifically, the reliance on conventional convolutional kernels limits the ability to capture complex nonlinear patterns and often leads to designs that lack interpretability, undermining their reliability in clinical applications.

The transformer-based model, i.e., SAM 2, does not achieve the best performance across the three regions, even with the application of point prompts. While the self-attention mechanism inherent to transformer architectures enables the contextual weighting of relevant information \cite{tuli2021convolutional}, this mechanism primarily enhances features that align with the global context. The suboptimal performance of SAM 2 on our dataset may be attributed to the nature of medical image segmentation, where features of interest often have limited relevance to broader contextual information. Research indicates that CNNs exhibit a stronger dependence on texture rather than shape when categorizing visual objects \cite{baker2018deep}, whereas transformers show the opposite tendency. The inclusion of a boundary loss function in our framework likely offsets CNNs' limitations in capturing object shape characteristics, which could contribute to the less favorable performance of SAM 2 in this evaluation.







%% file: conclusion.tex
\section{Conclusion}
We perform a comparative analysis of CNN-based architectures, i.e., U-Net, Att U-Net, and U-KAN, and a transformer-based segmentation network, i.e., SAM 2 for femur segmentation in MRI scan. The results reveal that CNN-based architectures generally outperform the transformer-based model in segmentation accuracy. However, the standalone U-Net model exhibits limited robustness, which is substantially enhanced by integrating attention mechanisms or the KAN framework, resulting in improved feature extraction and representation of bone structures. Among the models, Att U-Net achieves the highest precision, while U-KAN demonstrates the potential for enhanced predictive performance with the availability of larger datasets. These findings provide valuable insights into critical image features and model design considerations for accurate femur segmentation.
